\documentstyle[preprint,prd,aps]{revtex}

\newcommand{\sect}[1]{\setcounter{equation}{0}\section{#1}}

\newcommand{\bea}{\begin{eqnarray}}
\newcommand{\ena}{\end{eqnarray}}
\newcommand{\vs}[1]{\vspace{#1 mm}}

\renewcommand{\a}{\alpha}

\newcommand{\e}{\epsilon}
\newcommand{\dsl}{\pa \kern-0.5em /}

\newcommand{\pa}{\partial}

\begin{document}
\topmargin 0pt
\oddsidemargin 0mm

\renewcommand{\thefootnote}{\fnsymbol{footnote}}
\begin{titlepage}
\begin{flushright}
OU-HET 371\\
 MCTP-00-15\\
hep-th/0012239
\end{flushright}

\vs{5}
\begin{center}
{\large \bf The Galilean Nature of V-duality for Noncommutative 
Open String and Yang-Mills Theories}
\vs{5}

{\large Rong-Gen Cai\footnote{cai@het.phys.sci.osaka-u.ac.jp},
        J. X. Lu\footnote{jxlu@umich.edu},
and Yong-Shi Wu\footnote{wu@physics.utah.edu}\\} 

\vspace{4mm}

{\em $^\ast$ Department of Physics, Osaka University,
 Toyonaka, Osaka 560-0043, Japan\\
$^\dagger$ Michigan Center for Theoretical Physics, 
Randall Physics Laboratory\\ University of 
 Michigan,
 Ann Arbor MI 48109-1120, USA\\
$^\ddagger$ Department of Physics, University of Utah,
Salt Lake City, Utah 84112, USA\\}

\end{center}

\vs{5}
\centerline{{\bf{Abstract}}}
\vs{5}
\begin{small}
A V-duality conjecture for noncommutative open string 
theories (NCOS) that result from decoupling D-branes 
in Lorentz-boost related  backgrounds was put forward 
recently in hep-th/0006013. The aim of this paper is 
to test the Galilean nature of this conjecture in 
the gravity dual setup. We start with an (F, D3) bound 
state Lorentz-boosted along one D3-brane direction 
perpendicular to the F-string, and show that 
insisting a decoupled NCOS allows only infinitesimal 
Lorentz boosts. In this way, it is shown that the 
V-duality relates a family of NCOS by Galileo 
boosts. Starting with a Lorentz-boosted (D1,D3) 
bound state, we show that a similar V-duality  
works for noncommutative Yang-Mills (NCYM) theories 
as well. In addition, we deduce by a holography 
argument that the running string tension, as a 
function of the energy scale, for NCOS (or NCYM) 
remains unchanged under V-duality. 
\end{small}

\end{titlepage}

\newpage
\renewcommand{\thefootnote}{\arabic{footnote}}
\setcounter{footnote}{0}
\setcounter{page}{2}

\sect{Introduction}

	There is a surge of interest recently in seeking 
dynamical theories without gravity in string/M theory. 
The motivation is two-fold. On one hand, we try to 
understand better the Standard Model physics using the 
knowledge of string/M theory. One good example is the 
AdS/CFT correspondence and its variations. On the other 
hand, we intend to collect more theoretical data for the 
eventual formulation of M-theory since these decoupled 
theories are part of the big M-theory and we have a 
better hand on them due to the absence of gravity.

	Among these decoupled theories, for the purpose 
of this paper, we will focus on the so-called 
noncommutative Yang-Mills theory (NCYM)
\footnote{An incomplete list of references is 
\cite{how,conds,douh,howone,seiw,hasi,malr,aliosj}.} 
and noncommutative open string theory (NCOS)
\footnote{An incomplete list of references for NCOS is 
\cite{seist,gopmms,barr,har,klem,chew,lursone,russj,reyu,caio,lurstwo}.
}. In the simplest context, a (1+p)-dimensional 
noncommutative Yang-Mills with space-space 
noncommutativity arises in the decoupling limit as the 
decoupled theory of Dp branes in a constant magnetic 
background (or constant B-field with only spatial 
components). However, a decoupled noncommutative 
field theory does not exist for Dp branes 
in a constant electric background. Instead, 
one can have a decoupled open string theory 
living on the Dp brane worldvolume with 
space-time noncommutativity in the critical electric 
field limit. This is consistent with the fact that 
a unitary field theory cannot exist with space-time 
noncommutativity \cite{seistone,gomm}. The basic 
picture here is that in the decoupling limit the near 
critical electric force stretches the open string 
ending on the Dp brane, and balances the usual string 
tension to end up with almost tensionless string 
confined on the brane\footnote{We can keep the 
tension for the NCOS fixed by scaling the original 
tension to infinity.}. 

     It is well-known that we have various dualities 
such as $S-$ and $T-$, or in general $U-$dualities 
in the big M-theory. Up to this point, these dualities 
are global discrete transformations which relate 
different but physically equivalent vacua in  M-theory.
These dualities are believed to be inherited in the 
so-called little m-theory without gravity which is 
obtained as a decoupled theory of M-theory. NCOS and 
NCYM are among this little m-theory. Because of the 
absence of gravity (or in general the closed strings) 
in these decoupled theories, one may wonder whether 
there exist some new global transformations\footnote{We 
exclude those transformations which belong to the 
symmetry group of the underlying theory}, also called 
dualities, which connect physically equivalent theories.  
It has been suggested in \cite{chew} that there exists 
a new spacetime duality relating different but physically equivalent 
 NCOS which result from 
decoupling D-branes in Lorentz-boost related backgrounds. This 
can be understood as follows: The open string metric 
and space-time (and/or space-space) noncommutativity 
are both dictated by constant electric (and/or magnetic) 
background on the D-brane worldvolume. Some background
configurations are related by the worldvolume Lorentz 
boosts. The corresponding decoupled worldvolume noncommutative 
theories\footnote{The nature 
of theory (either NCOS or NCYM) cannot be changed 
since the Lorentz boost leaves ${\bf E}^2-{\bf B}^2$ 
invariant.} are conjectured to be physically equivalent 
to each other, since the Lorentz boosts are expected 
to be a symmetry of the parent perturbative open string 
theory in the decoupling limits. In this way, 
some decoupled theories with different noncommutativities 
and/or (open-string) metrics can be mapped to each 
other. For example, by first applying a Lorentz 
boost on a purely electric background and then imposing the 
corresponding decoupling limit, one can obtain a (1 + p)-dimensional NCOS 
with both space-time and space-space 
noncommutativities as a decoupled theory of Dp 
branes with constant electric and magnetic backgrounds. 
This NCOS is related to the original NCOS with only space-time 
noncommutativity\cite{chew} by the new spacetime duality. 
This conjectured duality 
was named V-duality in ref.\cite{chew}, because it 
originates from boosts, characterized by a relative 
{\it v}elocity, and alphabetically it follows the 
existing $S$-, $T$- and $U$-dualities.


Naturally raised is the following question:    
What is the action of the V-duality on metric 
and noncommutativities? Namely, what is
the transformation that connects the decoupled
theories arising from decoupling Dp branes 
in backgrounds related by a Lorentz boost\footnote{
We consider only those Lorentz boosts which do not 
leave the background fields invariant.}?
It is not necessarily a Lorentz boost, because
the decoupling limit involves intriguing scalings
of the background and the (closed-string) metric.
For example, consider a NCOS theory resulting
from decoupling D-branes in a background which is related
to a purely electric background by a Lorentz-boost. 
In ref. \cite{chew} an analysis using Seiberg-Witten 
relations \cite{seiw} indicated that the result
is a Galilean boost for the Lorentz boost in a direction 
orthogonal to that of the electric background. 
This is a particular case with orthogonal electric 
and magnetic backgrounds. 

In summary, the V-duality can be stated clearly 
in two parts: a) the NCOS's (or NCYM's) resulting 
from decoupling D-branes in Lorentz-related 
backgrounds are related to each other by Galileo 
boosts;\footnote{Recall that the noncommutative 
geometry for either NCOS or NCYM is described by 
the (open string) metric and the (anti-symmetric) 
coordinate noncommutative matrix.} 
and b) such related NCOS or NCYM theories are physically 
equivalent to each other. In this paper, we will test 
the V-duality from the gravity dual description of
NCOS, and will find that indeed the V-duality
holds exactly as suggested in \cite{chew}. Further,
we will extend this analysis of V-duality to NCYM
theories that result from decouplings of Dp branes 
in backgrounds Lorentz-boost related to a purely 
magnetic background and show that the V-duality 
holds for NCYM as well. For concreteness, we will 
focus on the case of $p = 3$ from now on, 
but the conclusion drawn is general.



In this paper, we limit ourselves to the particular 
cases in which the backgrounds are Lorentz-boost 
related to a purely electric or a purely magnetic 
background. However, our investigation indicates
that the V-duality holds in general, for example,
for NCOS and NCYM resulting from their respective
decouplings of D3 branes with backgrounds Lorentz-boost related
to a given parallel electric and magnetic background
as discussed in \cite{chew,lursone,lurstwo}. We plan
to report the general discussion of V-duality and its
structure, significance and relations to other
theories such as Matrix string theory in a separate
paper \cite{cailw}.
   
	This paper is organized as follows. In section 2, 
we first present a gravity configuration of (F, D3) 
bound state boosted along a brane direction orthogonal 
to the F-strings. This system provides the gravity
dual description of NCOS resulting from the decoupling 
of D3 branes in an orthogonal electric and magnetic
background (time-like). We show that the gravity
description is related to that of NCOS resulting from 
a purely electric background by a Galilean transformation. 
In other words, we show that the V-duality proposed 
in \cite{chew} holds also true in the gravity description. 
In section 3, we present a gravity configuration of 
(D1, D3) bound state boosted along a brane direction 
orthogonal to the D-strings. This system provides 
a gravity dual description of NCYM resulting from the 
decoupling of D3 branes in an orthogonal electric and
magnetic background (space-like). We show that similar 
V-duality holds here almost trivially. In section 4, 
we give a detailed explanation of holographic 
correspondence proposed in \cite{liw,chew} for NCYM 
and NCOS. Then we give a holographic derivation of 
the NSNS fields in the gravity dual description of 
NCOS discussed in section 2. We conclude this paper 
in section 5 where some issues on the validity 
of V-duality are also addressed.

\sect{The Galilean Nature of NCOS: V-duality}

We test the V-duality for NCOS using its gravity 
description in this section.
The relevant gravity configuration can be obtained by 
a Lorentz boost of the (F, D3) bound state 
\cite{rust,lurone} along a D3 brane direction
orthogonal to the F-strings in the bound state. 
The explicit form is
\begin{eqnarray}
&&d s^2 = H'^{- 1/2} \left[\frac{H'}{H} \left(- (d x^0)^2 
+ (d x^1)^2 + (d x^2)^2\right) \right.\nonumber\\
&&\qquad  + \left.\frac{4 \pi n g_s \a'^2}{r^4 \,H} 
\sin\a \tan\a \left(\cosh\gamma\, d x^2 - \sinh\gamma\, 
d x^0\right)^2 + (d x^3)^2\right] + H'^{1/2} \left[ 
d r^2 + r^2 d\Omega^2_5\right],\nonumber\\
&& e^\phi = g_s \sqrt{H'/H}, \ \ \ F_5 = 
16 \pi n \a'^2 (\ast \e_5 + \e_5), \nonumber\\
&&2\pi\a' B = \sin\a H^{- 1} \left(\cosh\gamma 
d x^0 \wedge d x^1 + \sinh\gamma d x^1 \wedge d x^2\right),
\nonumber\\
&& A_2 = g_s^{- 1} \tan\a H'^{- 1} \left(- \sinh\gamma 
d x^0\wedge d x^3 + \cosh\gamma d x^2\wedge d x^3\right),
\label{eq:one}
\end{eqnarray}
where we have the metric in string frame, $\e_5$ 
denotes the volume form of unit 5-sphere, $\ast$ 
denotes the Hodge dual in 10 dimensions, and the 
harmonic functions are
\begin{equation}
H = 1 + \frac{4\pi n g_s \a'^2}{r^4} 
\frac{1}{\cos\a},\ \ \ 
H' = 1 + \frac{4\pi n g_s \a'^2}{r^4} \cos\a.
\label{eq:onea}
\end{equation}
In the above, the integer $n$ is the number of D3 branes 
in the bound state and $g_s$ is the asymptotic string 
coupling. The boost is along the $x^2$ direction with boost 
parameter $\gamma$. When $\gamma = 0$, we recover the 
(F, D3) bound state. We can read the electric and 
magnetic background fields from the asymptotic B-field 
(i.e., $r \to \infty$) from the above as $E_1 = 
B^\infty_{01} = \sin\a \cosh \gamma, \ B_3 = 
B^\infty_{12} = \sin\a \sinh\gamma$. For 
$\gamma = 0$, we have only electric field. We have 
here ${\bf E}^2 - {\bf B}^2 = \sin^2\a \ge 0$ and 
${\bf E}\cdot {\bf B} = 0$ which are invariant under 
a Lorentz transformation as is evident. For NCOS, 
we need $\sin\a \ne 0$, therefore ${\bf E}^2 - {\bf B}^2 > 0$, 
i.e., time-like as it should be. The gravity dual 
description can be obtained from the above with the
following decoupling limit:
\begin{eqnarray}
&& \a' = \a'_{\rm eff} \e, \ \ \cos\a 
= \e, \ \ \gamma = \tilde v \e, \ \ 
g_s = \frac{G_o^2}{\e},
\nonumber\\
&& r = \a'_{\rm eff} \e^{1/2} u, \ \ 
x^{0, 1} = \frac{\tilde x^{0,
1}}{\sqrt{\e}}, x^{2, 3} = \sqrt{\e}\, \tilde x^{2,3},
\label{eq:two}
\end{eqnarray}
where $\e \to 0$ while parameters $\a'_{\rm eff}, 
\tilde v, G_o, u, \tilde x^\mu$ with ($\mu = 0, 1, 2, 3$) 
remain fixed. We would like to point out that insisting 
a decoupled gravity dual of  NCOS allows only an 
infinitesimal boost $\gamma = \tilde v \e$. This in 
turn says that a finite boost will not give a decoupled
NCOS in the decoupling limit. Therefore, a Lorentz boost 
on a given NCOS will render it undecoupled. This is 
entirely consistent with the conclusion drawn in 
\cite{chew} in a different analysis. With the above, 
the gravity description is then
\begin{eqnarray}
&&d s^2 = \e h^{1/2}\left[\frac{u^2}{R^2}\left(- 
d\tilde x_0^2 + d\tilde x_1^2 + h^{- 1} \left( 
d \tilde x_3^2 + \left( d \tilde x^2 - \tilde v 
d \tilde x^0\right)^2\right)\right) + \a'^2_{\rm eff} 
R^2 \left(\frac{d u^2}{u^2} + d\Omega_5^2\right)
\right],\nonumber\\
&&e^\phi = G_o^2 h^{1/2}, \ \ F_{0123 u} = 
-\e^2 \frac{16 \pi n u^3}{\a'^2_{\rm eff} R^8 h},
\nonumber\\
&& 2\pi \a' B = \e \frac{u^4}{R^4}  d \tilde x^0 
\wedge d\tilde x^1,\nonumber\\
&& A_2 = \e \frac{u^4}{G_o^2 R^4 h} \left(d\tilde x^2 
- \tilde v d \tilde x^0\right)\wedge d\tilde x^3,
\label{eq:three}
\end{eqnarray}
where 
\begin{equation}
h = 1 + \frac{u^4}{R^4},\ \ \ R^4 = 
\frac{4\pi n G_o^2}{\a'^2_{\rm eff}}.
\label{eq:threea}
\end{equation}

It is not difficult to examine that the parameter 
$\tilde v$ looks like a velocity and this gravity 
description is related to that of NCOS with only
space-time noncommutativity (corresponding to 
$\tilde v = 0$ or resulting from the background 
without boost) by a Galilean transformation
\begin{equation}
\tilde x^0 \to \tilde x^0, \ \ \ \tilde x^2 
\to \tilde x^2 - \tilde v \tilde x^0.
\label{eq:four}
\end{equation}
Further, if we calculate the noncommutative 
parameters, we have
\begin{equation}
\Theta^{01} = 2\pi \a'_{\rm eff}, \ \ \ 
\Theta^{12} = - 2\pi \alpha'_{\rm eff} \tilde v,
\label{eq:five}
\end{equation}
which is consistent with the Galilean transformation 
given in Eq. (\ref{eq:four}) above. Therefore, we in 
a gravity setup show that the V-duality conjecture 
proposed in \cite{chew} holds true.

   
\sect{The Galilean Nature of NCYM: V-duality}

The field theory analysis of NCYM for this case has 
been given in \cite{chew}. It concluded there that 
one cannot have NCYM with space-time noncommutativity 
in addition to the space-space one. This is also 
confirmed in other independent analysis, for example, 
in \cite{lursone,lurstwo}. This clearly indicates that 
one cannot perform a Lorentz transformation on NCYM 
since it will in general lead to space-time 
noncommutativity. Any allowed transformation on the 
decoupled theory must be induced from the transformation 
of string theory before the decoupling. We will show in 
this section that the allowed one is again a Galilean 
transformation, describing the V-duality action. This 
is entirely consistent with the fact that we cannot 
have space-time noncommutativity for NCYM since a 
Galilean transformation on a space-space noncommutativity 
cannot give rise to a space-time one. 

	In this section, we will spell out the V-duality 
action for NCYM using its gravity dual description. The 
relevant gravity configuration is the (D1, D3) bound 
state \cite{rust,bremm} boosted along a D3 brane
direction perpendicular to the D-strings in the 
state\footnote{This boosted configuration was also given 
in \cite{alior} for the discussion of light-like NCYM.}. 
Its explicit form is
\begin{eqnarray}
&&d s^2 = H^{- 1/2} \left[- (d x^0)^2 + (d x^1)^2 
+ (d x^2)^2 + \frac{H}{H'} (d x^3)^2\right.\nonumber\\
&&\qquad\qquad + \left. \frac{H - H'}{H'} 
(\cosh\gamma d x^2 - \sinh\gamma d x^0)^2 \right]
+ H^{1/2} \left[d r^2 + r^2 d\Omega_5^2\right],
\nonumber\\
&& e^\phi = g_s \sqrt{H/H'}, \ \ F_5 = 16 \pi 
n \a'^2 (\ast \e_5 + \e_5),\nonumber\\
&& 2\pi \a' B = H'^{- 1} \tan\a \left[\sinh\gamma 
d x^0 \wedge d x^3 -\cosh\gamma d x^2\wedge d x^3\right],
\nonumber\\
&& A_2 = g_s^{-1} H^{- 1} \sin\a \left[\cosh\gamma 
d x^0\wedge d x^1 + \sinh\gamma d x^1 \wedge d x^2\right]
\label{eq:six}
\end{eqnarray}
where $d s^2$ is the string metric and the harmonic 
functions $H$ and $H'$ are also given by Eq. 
(\ref{eq:onea}). We can read the D3 brane worldvolume 
electric and magnetic background fields from the 
asymptotic B-field as $E_3 = B^\infty_{03} = \tan\a 
\sinh\gamma, B_1 = B^\infty_{23} = - \tan\a \cosh\gamma$. 
Now we have ${\bf E}^2 - {\bf B}^2 = - \tan^2\a \le 0$ 
and ${\bf E}\cdot {\bf B} = 0$ which are Lorentz invariant. 
Since we are interested in NCYM, $\tan\a$ is nonzero. 
So ${\bf E}^2 - {\bf B}^2 < 0$, i.e., space-like as it 
should be for NCYM. The gravity dual of NCYM can be 
obtained from the above with the following decoupling 
limit:
\begin{eqnarray}
&&\a' = \a'_{\rm eff} \e, \ \ \cos\a = \e, \ \ 
\gamma = \tilde v \e, \ \ g_s = \e G_o^2,\nonumber\\
&&r = \a'_{\rm eff} \e u, \ \ x^{0, 1} = \tilde 
x^{0, 1}, \ \ x^{2, 3} = \e \tilde x^{2, 3},
\label{eq:seven}
\end{eqnarray}
where $\e \to 0$ while parameters $\a'_{\rm eff}, 
\tilde v, G_o, \tilde x^\mu$ with ($\mu = 0, 1, 2, 3$) 
remain fixed.  With the above decoupling limit, the 
gravity dual of NCYM is
\begin{eqnarray}
&&d s^2/\e = \frac{u^2}{R^2}\left[- d\tilde x_0^2 
+ d \tilde x_1^2 + h^{- 1} \left( d\tilde x_3^2 
+ \left(d \tilde x^2 - \tilde v \tilde d \tilde 
x^0\right)^2 \right)\right]\nonumber\\
&& \qquad\qquad + \a'^2_{\rm eff} R^2\left(
\frac{d u^2}{u^2} + d \Omega_5^2 \right)\nonumber\\
&&e^\phi = G_o^2 h^{- 1/2},\ \ F_{0123u} = 
-\e^2 \frac{16 \pi n u^3} {\a'^2_{\rm eff} R^8 h},
\nonumber\\
&&2\pi \a' B = - \e \frac{u^4}{R^4 h} \left(
d\tilde x^2 - \tilde v d\tilde x^0\right)\wedge 
d\tilde x^3,\nonumber\\
&&A_2 = \e \frac{u^4}{G_o^2 R^4} d\tilde x^0
\wedge d\tilde x^1, 
\label{eq:eight}
\end{eqnarray}
where  $h$ and $R$ are defined the same as those 
in Eq. (\ref{eq:threea}). We would like to point 
out again that the above decoupled gravity dual 
description of NCYM requires infinitesimal small 
boost parameter $\gamma = \tilde v \e$. A finite 
boost would imply that we cannot have the decoupled 
theory. This further implies that a Lorentz boost 
acting on a given decoupled NCYM will render it 
undecoupled.  

     One can examine that the above gravity dual of 
NCYM is related to that of NCYM resulting from the 
decoupling in a purely magnetic background (corresponding 
to $\tilde v = 0$) again by a Galilean transformation
defined in Eq. (\ref{eq:four}). If we calculate the 
noncommutativity parameters, we have, as expected,
\begin{equation}
\Theta^{0i} = 0, \ \ \ \Theta^{23} = 2\pi\a'_{\rm eff},
\end{equation}
where $i = 1, 2, 3$. This result is consistent with 
the Galilean transformation Eq. (\ref{eq:four}) as 
discussed at the outset of this section. So we show 
that V-duality also works for NCYM.

	We are not surprised by the infinitesimal boost 
requirement for NCYM given that for NCOS discussed in 
the previous section. This is because the two are S-dual 
to each other which can be explicitly checked from the 
gravity description for NCYM given in this section and 
that for NCOS in the previous section. So the V-duality 
here is also a consequence of S-duality. However, the 
physical reason behind is not explained up to this point. 
We try to provide this before turning to the next section. 

As we know that to have a decoupled NCOS, we need to 
have a background electric field present from the world
volume view. This electric field represents the presence 
of F-strings inside D3 branes. The stable BPS configuration
is the so-called non-threshold bound state of (F, D3). 
On the other hand, for NCYM we need to have magnetic 
background which represents the presence of D-strings 
inside D3 branes such that they form stable BPS 
non-threshold (D1, D3) bound state. So for D3 branes, the presence 
of D-strings favors the decoupled NCYM while that of F-strings 
favors the decoupled NCOS. We know that the decoupling limit 
for one against that for other. One cannot form a consistent 
hybrid theory which contains both NCOS and NCYM while at the 
same time it decouples the bulk closed strings. The effect 
of a Lorentz boost, for example, acting on (F, D3) is to 
create D-strings orthogonal to the F-strings in the bound 
state with its charge density proportional to the Lorentz 
boost. Given the previous discussion, we know that if the 
charge density for D-strings is comparable to that of 
F-strings in the bound state, we cannot have a decoupled 
NCOS\footnote{The best we can do is to infinitely boost a
vanishing small electric background to end up with a new 
light-like NCYM \cite{ahagm,alior}. The same applies for 
infinitely boosting a vanishing small magnetic background. 
As discussed in \cite{ahagm,alior}, the light-like NCYM, 
even though a noncommutative field theory, is qualitatively 
different from the usual NCYM with space-like 
noncommutativity, therefore a new kind of noncommutative 
field theory. One can say that NCOS, light-like NCYM and 
space-like NCYM result respectively from the decoupling 
of D3 branes in time-like, light-like and space-like 
electric and magnetic backgrounds. We stress that the
aforementioned infinite boost is used just for defining 
a finite light-like background from a vanishing either 
time-like or space-like one so that a decoupling limit 
for light-like NCYM can be addressed. This infinite boost 
is not directly related to V-duality. We will address the 
related V-duality issue in \cite{cailw}.}. Only for 
infinitesimally small D-string charge, a decoupled NCOS 
has a chance to exist. This is precisely what we achieved 
in the previous section and this is the physical reason 
why only a Galilean transformation or V-duality is allowed 
for the decoupled theory. The discussion for NCYM follows 
the same line but now the F-string charge density needs to 
be infinitesimal small, implying an infinitesimal boost. 
  

\sect{Holography and NCOS}

A holographic correspondence between NCYM and its 
gravity dual was proposed in \cite{liw}. This was 
further extended in \cite{chew} to NCOS with pure 
time-space noncommutativity. It basically states 
that the radial profile of the on-shell closed 
string moduli (string metric, NSNS B-field and 
dilaton) in the dual gravity description of NCYM 
or NCOS can be derived using Seiberg-Witten relations 
\cite{seiw} between fixed open string moduli 
(effective flat open string metric, noncommutative 
parameters and open string coupling) and the closed 
string moduli, provided a simple ansatz for the 
running string tension as the function of energy 
scale is assumed. Given V-duality, the holographic 
correspondence \cite{liw,chew} is expected to hold 
for NCOS and NCYM discussed in this paper as well. 
In this section, we provide a direct demonstration 
of this correspondence for the case of NCOS as an 
independent and direct check for V-duality. The 
check for NCYM is even easier and we will not 
repeat it here.

         Physically the holographic correspondence 
proposed in \cite{liw} for NCYM and in \cite{chew} 
for NCOS is a natural consequence of D-brane picture 
in the decoupling limit. We know that there are two 
physically equivalent descriptions of D-branes, one
is that of the open string ending on the D-branes 
and the other is the closed string one. 
  In general, from the closed string 
perspective, D-branes themselves interact with bulk closed string
modes and from the open string perspective, the open string ending on
the D-branes interacts with the same bulk modes. However, in the
so-called decoupling limit, these two descriptions describe the dynamics
of the same D-branes, therefore they must be physically equivalent. From
the open string perspective, the worldvolume must be flat since the bulk 
gravity (or closed string) decouples. However, the closed string 
description of
D-branes (i.e., the solitonic profile of D-branes in the decoupling
limit) gives a curved background (loosely called the near-horizon
geometry of D-brane\footnote{The original spacetime is separated to two 
regions: one is occupied by the bulk closed strings 
and the other describes the D-branes. The bulk closed 
strings cannot enter into the region of D-branes 
since its size is substringy. This is the picture 
of the decoupling of D-branes from the bulk closed 
strings in terms of the closed string (or bulk) description.}). 
It is the decoupled theory in a flat worldvolume obtained from
the open string perspective that gives a holographic 
description of the above curved background.

A better description for a decoupled theory can be obtained 
 when there are electric 
and/or magnetic background (or NSNS B-background) 
on the D-branes. In terms of this description, 
 the D-brane metric 
seen by the end-points of the effective open strings does not 
coincide with the closed string metric. And this description
gives rise to coordinate noncommutativity. In \cite{seiw} 
Seiberg and Witten have established the general 
relations between closed string moduli and the 
effective open string ones. 
 In general, these relations hold only to 
the leading order since a flat rigid geometry for 
D-branes is assumed there. However, in the decoupling 
limit for D-branes, these relations become exact 
since the bulk modes, which render the D-brane world
volume curved, decouple. Originally Seiberg-Witten 
relations are suggested in the first description of 
the D-branes (perturbative open-strings ending on 
flat D-branes). The important recognition made in 
\cite{liw,chew} is that these relations apply also 
to the closed string moduli in the dual gravity 
description. This provides a better understanding 
of how the gravity dual describes noncommutative 
theories in a holographical way. We can understand 
this as follows.

The open string description is non-gravitational 
(since the bulk closed strings decouple). Therefore 
the worldvolume geometry should be independent of 
energy scale, so are the metric and the 
noncommutativity parameters\footnote{One may wonder 
if the noncommutative parameters can be so, since 
they are dimensionful. The correct derivations given 
in \cite{liw,chew} ensure this.}. As we know, the 
energy scale in the open string description is just 
the radial coordinate $u$ in the closed string 
description of D-branes. So for a fixed $u$, we can 
ask for the effective description of open string. 
This effective open string coupling should be the 
same as before, since it is dimensionless, but the 
effective string tension should be determined properly. 
We therefore expect that Seiberg-Witten relations 
should hold between the closed string moduli at a fixed 
but arbitrary $u$ and the same open string moduli, but 
now for the effective open string with its yet unknown 
effective open string tension. The success of this 
prescription depends crucially on how to determine 
the effective string tension or $\a'_{\rm run}$. Given that the 
closed string moduli in original Seiberg-Witten relations are 
constants and independent of the $u$ coordinate, they must correspond 
to those outside of the boundary of the gravity description.
This fact helps us to set the boundary condition for $\a'_{\rm run}$
which was also discussed in \cite{liw,chew} in a different context.
In other words,  it must approach to its boundary 
value as $u \to \infty$.

	In the following, we derive the NSNS fields 
for the gravity dual of NCOS given in Eq. 
(\ref{eq:three}) as an example. The Seiberg-Witten 
relations \cite{seiw} between open string moduli 
($G_{\mu\nu}, \Theta^{\mu\nu}, G_o$) and closed string 
ones ($g_{\mu\nu}, B_{\mu\nu},g_s$) in their original 
forms are
\begin{equation}
G_{\mu\nu} = g_{\mu\nu} - (2\pi \a')^2 
(B g^{- 1} B)_{\mu\nu},
\label{eq:ninea}
\end{equation}
\begin{equation}
\Theta^{\mu\nu} = 2 \pi \a' \left(\frac{1}{g 
+ 2\pi \a' B}\right)^{\mu\nu}_A,
\label{eq:nineb}
\end{equation}
and
\begin{equation}
G_s = G_o^2 = g_s \left(\frac{\det G}
{\det(g + 2\pi \a' B)}\right)^{1/2}.
\label{eq:ninec}
\end{equation}
In the above, $(\,)_A$ denotes the anti-symmetric 
part and $\mu = 0, 1, 2, 3$.

The open string metric $G_{\mu\nu}$ and the 
noncommutative parameters $\Theta^{\mu\nu}$ 
can be calculated using the asymptotic values 
for closed  string metric and NSNS B-field from 
Eq. (\ref{eq:one}) with respect to the scaled 
coordinates $\tilde x^\mu$ as
\begin{equation}
G_{\mu\nu} = \e \left(\begin{array}{cccc}
                       - (1 - \tilde v^2)&0&-\tilde v&0\\
                       0&1&0&0\\
                       -\tilde v &0&1&0\\
                       0&0&0&1\\
                       \end{array}\right),
\label{eq:tena}
\end{equation}
and
\begin{equation}
\Theta^{01} = 2\pi\a'_{\rm eff},\ \ 
\Theta^{12} = - 2\pi\a'_{\rm eff}\tilde v.
\label{eq:tenb}
\end{equation}
We know that the open string metric $G_{\mu\nu}$ 
is defined with respect to the string tension 
$1/(2\pi \a')$. We can redefine a new open string 
metric $G'_{\mu\nu} = G_{\mu\nu}/\e$ with a
finite tension $1/(2\pi\a'_{\rm eff})$ since 
$\e = \a'/\a'_{\rm eff}$. We now have $\a'G^{\mu\nu} 
= \a'_{\rm eff} G'^{\mu\nu} = {\rm fixed}$ as 
expected. The new metric $G'_{\mu\nu}$ can be 
obtained from Eq. (\ref{eq:ninea}) simply by 
setting closed string metric $g \to g/\e$ and 
$\a' \to \a'_{\rm eff}$. In other words, we now 
work with the metric $d s^2/\e$ and string constant 
$\a'_{\rm eff}$. So $\a'_{\rm eff}$ is the boundary 
value for the $\a'_{\rm run}$. The question is how 
to obtain a correct ansatz for $\a'_{\rm run}$. 

Consider a type IIB F-string probe in a given
background, its kinetic term is  
\begin{equation}
\frac{1}{2 \pi \a'}\int d^2\sigma 
\partial_\a x^M \partial^\a x^N g_{MN},
\end{equation}
with $g_{MN}$ the bulk metric. Now consider this 
string to probe the D3 brane background. Recall 
that the NCOS tension is the same with respect 
to its metric in any direction. But the story is 
different if we try to read the running tension 
from the closed string side. The reason that we 
have a decoupled NCOS is because the near-critical 
electric  force balances the original string tension 
to end up with a finite tension if the original
tension is sent to infinity. This NCOS is now  
along $x^1$ direction. It is in this direction that 
the open string metric is different from the 
asymptotic closed string one. One cannot use the 
probe F-string to read the running tension for 
the effective open string along this direction. 
Examining the open string metric $G_{\mu\nu}$ given 
in Eq. (\ref{eq:tena}), we can see that the open
string metric $G_{22}, G_{33}$ are the same as 
the corresponding asymptotic closed metric $g_{22}, 
g_{33}$ with respect to the scaled coordinates 
$\tilde x^\mu$. We should be able to read the 
running tension for the effective open string 
using the probe along, for example, $\tilde x^3$ 
direction. In other words, from $1/(2\pi \a') \int
\partial_\a \tilde x^3 \partial^\a \tilde x^3 
g_{33} (u) = g_{33} (u) /(2\pi \a') \int \partial_\a 
\tilde x^3 \partial^\a \tilde x^3 $, we have 
$\a'_{\rm run} = \a'/g_{33}$. From Eq. (\ref{eq:three}), 
we have $\a'_{\rm run} = \a'_{\rm eff} R^2 h^{1/2} /u^2$. 
It is easy to see that $\a'_{\rm run} \to \a'_{\rm eff}$
as $u\to \infty$, i.e., satisfying the boundary 
condition as expected. The physical picture above 
is: The closed string has a string constant $\a'$ but 
moving in a curved background, for example, $g_{33}$ 
(or $\a'_{\rm eff}$ but in $g_{33}/\e$) while the 
open string has a flat metric $G_{\mu\nu}$ but with 
a running $\e \a'_{\rm run}$ (or $G'_{\mu\nu}$ with 
$\a'_{\rm run}$).

With the above understanding and given 
\begin{equation}
\a'_{\rm run} = \a'_{\rm eff} R^2 h^{1/2} /u^2, 
\label{alpharun}
\end{equation}
we now present the holographic derivation for the 
NSNS fields in the gravity dual of NCOS given in 
Eq. (\ref{eq:three}). Given that the gravity dual 
of NCOS in (\ref{eq:three}) is obtained from a 
background with orthogonal electric and magnetic 
fields and the magnetic background is brought
about by a Lorentz boost, we expect that the only 
nonvanishing metric along the brane directions 
are $g_{00}, g_{11}, g_{02} = g_{20}, g_{22} = g_{33}$.  
We also have possible nonvanishing $B_{01}$ and
$B_{12}$.  In the following, we use open string 
metric $G'_{\mu\nu} = G_{\mu\nu}/\e$, this implies 
that we use closed string metric $g/\e$. Given the 
ansatz for $\a'_{\rm run}$, it implies that we know
$g_{22}/\e = g_{33}/\e = \a'_{\rm eff}/\a'_{\rm run}$ 
as is evident from the above discussion. Now the 
$\a'$ in Eqs. (\ref{eq:ninea}) -- (\ref{eq:ninec}) 
is replaced by $\a'_{\rm run}$. This further implies
that the closed string metric used in Seiberg-Witten 
relations should be $\bar g_{\mu\nu} = g_{\mu\nu}/g_{33}$. 
With the above in mind, we set $\bar g_{00} = 
- f_1, \bar g_{11} = f_2, \bar g_{02} = f_3$ and
$\bar g_{22} = \bar g_{33} = 1$. For simplicity, we 
set $2\pi \a'_{\rm run} B_{01} = h_1, 2\pi\a'_{\rm run} 
B_{12} = h_2$. Then Seiberg-Witten relations give, 
from the metric,
\begin{equation}
1 + \frac{h_2^2}{f_2} = 1, \ \ f_3 
= - \tilde v, \ \ f_1 -\frac{h_1^2}{f_2} 
= 1 - \tilde v^2, \ \ f_2 - \frac{h_1^2}
{f_1 + f_3^2} = 1,
\label{eq:elevena}
\end{equation}
and, from the noncommutative matrix,
\begin{equation}
\frac{h_1}{h_1^2 - f_2 (f_1 + f_3^2)} 
= - \frac{u^2}{R^2 h^{1/2}}.
\label{eq:elevenb}
\end{equation}
One can solve from the above to have 
\begin{equation}
f_1 = h - \tilde v^2, \ \ f_2 = h, \ \ 
f_3 = - \tilde v, \ \ h_1 =
\frac{u^2}{R^2} h^{1/2}, \ \ h_2 = 0.
\label{eq:twelve}
\end{equation}

	One can check that these solutions give
the correct metric and NSNS B-field in Eq.
(\ref{eq:three}). For example, we have $2\pi
\a'_{\rm run} B_{01} = h_1$ which gives 
$ 2\pi \a'_{\rm eff} B_{01} = u^4/R^4$, the 
correct answer. Using the above results, we 
can calculate the effective closed string 
coupling $e^\phi$, from $e^\phi = G_o^2 [- \det 
(\bar g + 2\pi \a'_{\rm run} B)]^{1/2}$, as
\begin{equation}
e^\phi = G_o^2 h^{1/2},
\end{equation}
again the correct answer. 
We note that the running string tension for NCOS 
in our case, Eq. (\ref{alpharun}), as a function 
of the radial coordinate $u$ is the same as that 
given in ref. \cite{chew} for NCOS with pure 
space-time noncommutativity. This means that the
NCOS's related by V-duality have the same running 
string tension. Or the latter remains unchanged
under V-duality. This may be viewed as an additional
evidence for V-duality.


\sect{Conclusions and Discussions}

Before concluding this paper, let us now address an issue
regarding the supporting argument for the equivalence of
the noncommutative theories that arise from decoupling
D-branes in Lorentz-boost related backgrounds. This issue arises as 
follows: In the usual perturbative open string 
description of D-branes, from the outset, one assumes that 
D-brane are rigid and flat. This is known to be not exactly 
true, since the coupling to bulk closed strings will 
render the D-branes curved. In this sense, the Poincare 
symmetry along the brane directions in this description 
should be considered true only approximately. However, in a limit 
in which the bulk closed strings decouple, this symmetry 
is expected to become exact. Moreover, the decoupling for a
noncommutative theory (either NCOS or NCYM for the purpose 
of this paper) in general requires a D-brane worldvolume  
electric and/or magnetic background field. The presence of 
such a background field breaks the worldvolume Lorentz 
symmetry, but this breakdown is only spontaneous. 
Namely, 
if we simultaneously Lorentz transform all backgrounds,
including the electromagnetic one, we end up with a
physically equivalent situation. This can be explicitly
seen in our present study: in the gravity dual 
description in Sec. 2 and 3 it is obvious that the Lorentz 
boost on a relevant D-brane bound state solution before 
decoupling leads to a physically equivalent solution. 
Thus, given that the worldvolume Lorentz symmetry for the 
parent open string theory should be exact in decoupling 
limits and that the introduction of backgrounds breaks 
the symmetry only spontaneously, we expect that the 
noncommutative theories of Dp branes resulting from 
decoupling in Lorentz-boost related background fields 
are physically equivalent.

In conclusion,  we test the V-duality conjecture proposed 
in \cite{chew} for NCOS using its gravity dual description.  
We also extend the V-duality for NCYM and find that this 
seems to be true in general. The implication of this 
V-duality is that one can no longer use a Lorentz 
transformation other than a Galilean one, for example, 
to choose a particular frame in performing calculations 
in NCOS and NCYM.
We present an explanation for the holographic 
correspondence given in \cite{liw,chew} for 
noncommutative theories, and give a holographic 
derivation for the gravity dual of NCOS discussed in 
this paper. In addition, our holographic derivation 
shows that the NCOS's (or NCYM's) related by V-duality 
share the same running string tension as a function
of the radial coordinate (or energy scale).   

The Galilean nature of this V-duality is, in a sense,  
implied already by the noncommutative relation 
$[x^\mu, x^\nu] = i \Theta^{\mu\nu}$, which implies
spacetime nonlocality. This in turn implies
that we have action at distance. In other words, we do not have a limit
on the speed of light\footnote{There are some subtleties involved when
the boost is not orthogonal to the space-time noncommutative
directions. We will address this in detail in \cite{cailw}.}, 
i.e., we are dealing with Newtonian-like physics.
With the V-duality discussed in this paper and the above simple physical 
implication behind the noncommutative relations, one can easily 
understand the recent finding in
\cite{hast,leep} that solitons can travel faster than the speed of light 
in NCYM and also the infinitely large speed of light as discussed in
\cite{dangkone,dangktwo,gomo}. We will elaborate this and other related 
points further in the forthcoming paper \cite{cailw}. 

\section*{Acknowledgements}
The work of RGC was supported in part by the Japan 
Society for Promotion of Science and in part by 
Grants-in-Aid for Scientific Research Nos. 99020 
and 12640270, and by a Grant-in-Aid on the Priority 
Area: Supersymmetry and Unified Theory of Elementary 
Particles. JXL acknowledges the support of U. S. 
Department of Energy. The research of YSW was 
supported in part by the U.S. National Science 
Foundation under Grant No. PHY-9907701.

\end{document}